\documentclass[11pt]{article}
\usepackage{amsmath}
\usepackage{graphicx}
\newtheorem{remark}{Remark}
\def\re{\mbox{Re\;}}
\def\sech{\mbox{sech\;}}

\def\const{\mbox{const\;}}
\def\im{\mbox{Im\;}}
\bigskip

\begin{document}
\title{Asymptotic Behavior of Manakov Solitons: Effects of
Potential Wells and Humps}

\author{V. S. Gerdjikov\thanks{E-mail: gerjikov@inrne.bas.bg}, A. V. Kyuldjiev\thanks{E-mail: kyuljiev@inrne.bas.bg}\\
{\small\textit{Institute for Nuclear Research and Nuclear Energy, Bulgarian
Academy of Sciences,}}\\
{\small\textit{72 Tzarigradsko Chaussee, Blvd., 1784 Sofia, Bulgaria }}\\ \\
M. D. Todorov\thanks{E-mail: mtod@tu-sofia.bg}\\
{\small\textit{Department of Applied Mathematics and Computer Science, Technical University of Sofia,}}\\
{\small\textit{8 Kliment Ohridski, Blvd., 1000 Sofia, Bulgaria}}} 

\date{}

\maketitle

\begin{abstract}
We consider the asymptotic behavior of the soliton solutions of Manakov's system perturbed by external potentials. It has already been
established that its multisoliton interactions in the adiabatic approximation can be modeled by the Complex Toda chain (CTC). The fact that
the CTC is a completely integrable system, enables us to determine the asymptotic behavior of the multisoliton trains.
In the present study we accent on the 3-soliton initial configurations perturbed by sech-like external potentials and compare the
numerical predictions of the Manakov system and the perturbed CTC in different regimes. The results of conducted analysis show that
the perturbed CTC can reliably predict the long-time evolution of the Manakov system.                                                    \end{abstract}
\bigskip

                    \section{Introduction}

The Gross-Pitaevski (GP) equation  and its multicomponent generalizations are important tools
for analyzing and studying the dynamics of the Bose-Einstein condensates (BEC), see the monographs
\cite{38,25,29} and the numerous references therein among which we mention \cite{40,5,4,
27,39,49,19,34,26,8,30,32}.  In the 3-dimensional case these equations can be analyzed solely by numerical methods.
If we assume that BEC is quasi-one-dimensional then the GP equations mentioned above may be  reduced to the nonlinear Schr\"odinger  equation (NLSE)
perturbed by the external potential $V(x)$
\begin{equation}\label{eq:NLS0}\begin{split}
i u_t + \frac{1}{2} u_{xx} + |u|^2 u(x,t) =V(x)u(x,t),
\end{split}\end{equation}
or to its vector generalizations (VNLSE)
\begin{equation}\label{eq:vnls}\begin{split}
i\vec{u}_t + \frac{1}{2} \vec{u}_{xx} +(\vec{u}^\dag, \vec{u}) \vec{u}(x,t)= V(x)\vec{u}(x,t).
\end{split}\end{equation}
The Manakov model (MM) \cite{35} is a two-component  VNLSE with $V(x)=0$ (for more details see  \cite{1,24}).

The analytical approach to the $N$-soliton interactions was proposed by Zakharov and Shabat \cite{51,37}
for the scalar NLSE, for a vector NLSE see \cite{31}. They treated the case of the exact $N$-soliton solution where all solitons had different
velocities. They calculated the asymptotics of the $N$-soliton solution for $t\to \pm \infty$ and proved
that both asymptotics are sums of $N$ one-soliton solutions with the same sets of amplitudes and velocities.
The effects of the interaction were shifts in the relative center of masses and phases of the solitons.
The same approach, however, is not applicable to the MM, because the asymptotics of the soliton
solution for $t\to \pm\infty$ do not commute.

The present paper is an extension of \cite{12,20,21} where the main result is that the $N$-soliton
interactions in the adiabatic approximation for the Manakov model can also be modeled by the CTC
\cite{16,22,14,23}.
More specifically, here we continue our analysis of the effects of external potentials on the soliton interactions.
While in \cite{12,20,8} we studied the effects of periodic, harmonic and anharmonic potentials, here we
consider potential wells and humps of the form:
\begin{equation}\label{eq:iRu}\begin{split}
V(x)=\sum_{s}^{} c_s V_s(x,y_s),  \qquad   V_s(x,y_s)=\frac{1}{\cosh^2 (2\nu_0 x-y_s)}.
\end{split}\end{equation}
If  $c_s $ is negative (resp. positive) $V_s(x)$ is a well (resp. hump)
with width 1.7 at half-height/depth. Adjusting one or more terms in (\ref{eq:iRu})
with different  $c_s$ and $y_s$ we can describe wells and/or humps with different widths/depths and positions.

In the present paper we in fact prove the hypothesis in \cite{7} and extend the results in
\cite{7,8,11,33,43,50,12,10,17} concerning the model of soliton interactions of vector NLSE (\ref{eq:vnls})
in adiabatic approximation.

The corresponding vector $N$-soliton train is a solution of (\ref{eq:vnls})
determined by the initial condition:
\begin{equation}\label{eq:Nstr}\begin{split}
\vec{u}(x,t=0) &= \sum_{k=1}^N \vec{u}_k(x,t=0), \qquad \vec{u}_k(x,t) = u_k(x,t)
 \vec{n}_k , \qquad u_k(x,t) = {2\nu_k {\rm e}^{i\phi_k}\over \cosh(z_k)}
\end{split}\end{equation}
with
\begin{equation}\label{eq:Nstr2}\begin{aligned}
z_k &= 2\nu_k (x-\xi_k(t)), &\qquad \xi_k(t) &=2\mu_k t +\xi_{k,0}, \\
\phi_k &= {\mu_k \over \nu_k} z_k + \delta_k(t), &\qquad
\delta_k(t)&=2(\mu_k^2+\nu_k^2) t +\delta_{k,0},
\end{aligned}\end{equation}
where the  $s$-component polarization vector $\vec{n}_k = \left( n_{k,1} e^{i \beta_{k,1}}, n_{k,2} e^{i \beta_{k,2}} ,\dots ,
n_{k,s} e^{i \beta_{k,s}}\right)^T$ is normalized by the conditions
\begin{equation}\label{eq:n-k2}\begin{split}
\langle \vec{n}_k ^\dag , \vec{n}_k\rangle \equiv \sum_{p=1}^s n_{k,p}^{2}= 1, \qquad \sum_{p=1}^{s} \beta_{k;s} =0.
\end{split}\end{equation}
The adiabatic approximation holds true if the soliton parameters  satisfy \cite{28}:
\begin{eqnarray}\label{eq:ad-ap}
&& |\nu _k-\nu _0| \ll \nu _0, \qquad |\mu _k-\mu _0| \ll \mu _0,
\qquad |\nu _k-\nu _0| |\xi_{k+1,0}-\xi_{k,0}| \gg 1,
\end{eqnarray}
for all $k$, where $\nu _0 = {1  \over N }\sum_{k=1}^{N}\nu _k$, and $ \mu _0 =
{1 \over N }\sum_{k=1}^{N}\mu _k$ are the average amplitude and
velocity, respectively. In fact we have two different scales:
\[ |\nu _k-\nu _0| \simeq \varepsilon_0^{1/2}, \qquad |\mu _k-\mu _0| \simeq \varepsilon_0^{1/2}, \qquad
|\xi_{k+1,0}-\xi_{k,0}| \simeq \varepsilon_0^{-1}.\]

We remind that the basic idea of the adiabatic approximation is to derive a dynamical system for the soliton
parameters which would describe their interaction. This idea was initiated by Karpman and Solov'ev \cite{28} and
modified by Anderson and Lisak \cite{2}. Later this idea was generalized to $N$-soliton interactions
\cite{22,16,14,23} and the corresponding dynamical system for the $4N$-soliton parameters was identified as a
$N$-site complex Toda chain (CTC). The fact that the CTC, (just like its real counterpart -- the Toda chain) is completely
integrable gives additional possibilities. A detailed comparative analysis between the solutions of the RTC and CTC
\cite{13} shows that the CTC allows for a variety of asymptotic regimes, see Section 3 below. More precisely, knowing the
initial soliton parameters one can effectively predict the asymptotic regime of the soliton train. Another possible use
of the same fact is, that one can describe the sets of soliton parameters responsible for each of the asymptotic regimes.
Another important advantage of the adiabatic approach consists in the fact, that one may consider the effects of various
perturbations on the soliton interactions \cite{28,22}.

The next step was to extend this approach to treat the soliton interactions of the Manakov solitons.
More precisely, using the method of Anderson and Lisak \cite{2}
we derive a generalized version of the Complex Toda Chain (CTC)   (see Eqs. (\ref{eq:141.1}), (\ref{eq:q_k})
below) as a model describing the behavior of the $N$-soliton trains of of the VNLSE (\ref{eq:vnls})
\cite{7,8,10,12,21,17}. This generalized
CTC includes dependence  on the polarization vectors $\vec{n}_k$. It allows us to analyze how the changes of the polarization vectors
influence the soliton interactions. Besides, the generalized CTC is also integrable with the consequence that one can predict
the asymptotic regimes of the Manakov solitons and can describe the sets of soliton parameters that are responsible for
each of the asymptotic regimes. Of course, just like for the scalar case, one can also analyze the effects of the
various perturbations on the soliton interactions.


In Section 2 we  outline how the variational approach developed in \cite{2}
can be used to derive  the perturbed CTC (PCTC) model \cite{12,10} for $\sech$-type external potentials.
We also remind the reader about the asymptotic regimes of the soliton trains predicted by the CTC \cite{16,14}.
In Section 3 we briefly treat the $N$-soliton interactions of the MM without external
potential. Obviously in order to determine the $N$-soliton train for the MM, along with the usual sets of
solitons parameters $\nu_k,\mu_k,\xi_k$ and $\delta_k$ we need also the set of polarization vectors
$\vec{n}_s$.
In Section 4 we derive the effects of the external potentials on the soliton interactions.
This is a perturbed  form of the CTC  (PCTC) for generic potentials of the form (\ref{eq:iRu}).
Section 5 is dedicated to the comparison between the numeric solutions of the perturbed VNLSE
(\ref{eq:vnls}) with the predictions of the PCTC model.
 To this end we solve the VNLSE numerically by using an implicit scheme of  Crank-Nicolson type in complex arithmetic. The concept of the internal iterations is applied (see \cite{6}) in order to ensure the implementation of the conservation laws on difference level within the round-off error of the calculations \cite{45,46,47}.
The solutions of the relevant PCTC
have been obtained using Maple. Knowing the numeric solution $\vec{u}$ of the perturbed VNLSE we calculate  he maxima of $(\vec{u}\,^\dag, \vec{u})$, compare them with the (numeric solutions) for $\xi_k(t)$ of the PCTC and plot the presdicted by both models trajectories for each of the  solitons. Thus we are able to analyze the effects of the external potentials on the soliton interactions. Finally, Sections 6 and 7 contain discussion and conclusions.

\section{Preliminaries}

Here  we briefly remind the derivation of the CTC as a model describing the $N$-soliton interactions
VNLS systems using the variational approach \cite{2}.

\subsection{Derivation of the CTC as a model for the soliton interaction of perturbed VNLS systems}


The perturbed vector NLSE (\ref{eq:vnls}) allows Hamiltonian
formulations with the Poisson brackets
\begin{equation}\label{eq:PB}
    \{ \vec{u}_j(x,t), \vec{u}_k^*(y,t)\} = \delta_{jk} \delta(x-y)
\end{equation}
and the Hamiltonian
\begin{equation}\label{eq:120.1}
H=\int_{-\infty}^{\infty} dx\; \left[ {1\over 2} \langle\vec{u}_x,
\vec{u}_x\rangle - {1\over 2} \langle\vec{u}\,^\dag, \vec{u}\rangle^2 +V(x) \langle\vec{u}\,^\dag, \vec{u}\rangle \right].
\end{equation}
It also admits a Lagrangian of the form:
\begin{equation}\label{eq:120.2}
\mathcal{L}= \int_{-\infty}^{\infty} dt\;  {i\over 2} \left[
\langle\vec{u}, \vec{u}_t\rangle- \langle\vec{u}_t,\vec{u}) \right] - H.
\end{equation}

In what follows we will analyze the large time behavior of
the $N$-soliton train determined as the solution of the VNLSE by the initial condition (\ref{eq:Nstr}),  (\ref{eq:Nstr2}).

The idea of the variational approach of \cite{2} is to insert the anzatz (\ref{eq:Nstr}) into the Lagrangian,
perform the integration over $x$ and retain only terms of the orders of $\varepsilon_0^{1/2}$ and $\varepsilon_0$.
The first obvious observation is that only the nearest neighbors solitons will contribute such terms and
\begin{equation}\label{eq:L}\begin{split}
\mathcal{L} &= \sum_{k=1}^{N} \mathcal{L}_k + \sum_{k=1}^{N} \sum_{n=k\pm 1} \widetilde{\mathcal{L}}_{kn}.
\end{split}\end{equation}
Here $\mathcal{L}_k $ correspond to the terms involving only the $k$-th soliton (see \cite{12}):
\begin{equation}\label{eq:123.2}\begin{split}
\mathcal{L}_k = 4\nu _k \left( {i  \over 2 } \left( \langle\vec{n}_k,\vec{n}_{k,t}\rangle
-\langle\vec{n}_{k,t},\vec{n}_k\rangle \right) +2\mu _k{d\xi_k  \over dt } - {d\delta _k  \over dt }
-2\mu _k^2 + { 2\nu _k^2 \over 3 }\right) -\int_{x=-\infty}^{\infty} V(x) \langle\vec{u}_k\,^\dag, \vec{u}_k\rangle.
\end{split}\end{equation}
Finally the terms describing soliton-soliton interactions are given by:
\begin{equation}\label{eq:}\begin{split}
\mathcal{L}_{kn}&= 16\nu _0^3 e^{-\Delta_{kn}} (R_{kn}+R_{kn}^*)+\mathcal{O}(\epsilon ^{3/2}),\\
R_{kn}&= e^{i(\widetilde{\delta} _n- \widetilde{\delta} _k)} \langle\vec{n}_k, \vec{n}_n\rangle, \qquad
\widetilde{\delta}_k=\delta _k - 2\mu _0\xi_k, \; \Delta _{kn} =2s_{kn}\nu _0(\xi_k- \xi_n),
\end{split}\end{equation}
where $s_{k,k+1}= 1$ and $s_{k,k-1}= -1$.

The next step is to consider $\mathcal{L}$ (\ref{eq:L}) as a Lagrangian of the dynamical system,
describing the motion of the $N$-soliton train and providing the equations of motion for the
$(2s+2)N$ ($6N$ for the Manakov case) soliton parameters.

Let us first consider the unperturbed case, \emph{i.e.}, $V(x)=0$.
Deriving the dynamical system  we get terms of three different orders of magnitude: (i)~terms of order
$\Delta^2_{kn}\exp(-\Delta_{kn})$; (ii)~terms of order $\Delta_{kn}\exp(-\Delta_{kn})$ and (iii)~terms of order
$\exp(-\Delta_{kn})$. However the terms of types (i) and (ii) are multiplied by factors that are of
the order of $\exp(-\Delta_{kn})$ due to the evolution equations
for the soliton parameters. Finally,  we arrive at the following set
of dynamical equations for the soliton parameters:
\begin{equation}\label{eq:139.2}\begin{aligned}
{d\xi_k  \over dt } &= 2\mu _k , &\quad {d\delta _k  \over dt } &=  2\mu _k^2 +2\nu_k^2 , \\
{d\nu _k  \over dt } &= 8\nu _0^3 \sum_{n}^{} e^{-\Delta _{kn}} i(R_{kn}-R_{kn}^*), &\quad
{d\mu _k  \over dt } &= -8\nu _0^3 \sum_{n}^{} e^{-\Delta _{kn}} (R_{kn}+R_{kn}^*).
\end{aligned}\end{equation}
In addition we obtain also a system of equations for the evolution of the polarization vectors:
 \begin{equation}
\label{eq:139.6} {d\vec{n}_k  \over dt } = 4\nu _0^2 i\sum_{n=k\pm
1}^{} e^{-\Delta _{kn}}\left[e^{i(\widetilde{\delta }_n -
\widetilde{\delta }_k)} \vec{n}_n -R_{kn}\vec{n}_k  +
e^{i(\widetilde{\delta }_n - \widetilde{\delta }_k)} \vec{n}_n
+R_{kn}^*\vec{n}_k \right] + C_k\vec{n}_k .
\end{equation}
where the constants $C_k$ are fixed up by the constraints on the polarization vectors.
Indeed, from  $\langle\vec{n}_k\;^\dag,\vec{n}_k\rangle =1$ for all $t$ one finds that $C_k+C_k^*=0$,
\emph{i.e.}, the constants $C_k$ are purely imaginary. Let us now assume that $C_k =i \theta_k $.
Then from eqs. (\ref{eq:n-k2}) and (\ref{eq:139.6}) we find that $\beta_{k,s}$ become time-dependent
and up to terms of the order of $\epsilon$ evolve linearly with time: $\beta_{k,s}(t) =
\beta_{k,s}(0) +\theta_k t$. But such evolution is compatible with the second normalization
condition in (\ref{eq:n-k2}) only if $\theta_k=0$; therefore $C_k=0$.
Thus,  from Eqs. (\ref{eq:139.2})  we get:
\begin{equation}\label{eq:141.1}
{d(\mu _k+i\nu _k)  \over dt } = 4\nu _0 \left[
\langle\vec{n}_{k}, \vec{n}_{k-1}\rangle e^{q_{k}-q_{k-1}} -
\langle\vec{n}_{k+1}, \vec{n}_k\rangle e^{q_{k+1}-q_{k}} \right],
\end{equation}
where
\begin{equation}\label{eq:q_k}\begin{aligned}
q_k &= -2\nu _0\xi_k + k \ln 4\nu _0^2 - i (\delta _k+\delta _0 + k\pi
-2\mu _0 \xi_k), \\
\nu _0 &= {1 \over N } \sum_{s=1}^{N} \nu _s, \qquad
\mu _0 = {1 \over N } \sum_{s=1}^{N} \mu _s, \qquad
\delta _0 = {1 \over N } \sum_{s=1}^{N} \delta _s.
\end{aligned}\end{equation}

Besides, from (\ref{eq:139.2}) and (\ref{eq:q_k}) there follows (see \cite{22}):
\begin{equation}\label{eq:141.2}
{dq_k  \over dt } =-4\nu _0 (\mu _k + i\nu _k).
\end{equation}
and
\begin{equation}\label{eq:141.3}
{d^2q _k \over dt^2 } = 16\nu _0^2 \left[ \langle\vec{n}_{k+1}, \vec{n}_k\rangle e^{q_{k+1}-q_{k}} - \langle\vec{n}_{k},
\vec{n}_{k-1}\rangle e^{q_{k}-q_{k-1}} \right],
\end{equation}
which   proves the statement in \cite{7}. Eq. (\ref{eq:141.3}), combined with the system of equations for the
polarization vectors (\ref{eq:139.6}) provides the proper generalization of the CTC model for the MNLS.

The equations for the polarization vectors are nonlinear. So the
whole system of equations for $q_k$ and $\vec{n}_k$ seems to be rather complicated and
nonintegrable even for the unperturbed MNLS. However, all terms in the right hand sides of the
evolution equations for $\vec{n}_k$ are of the order of
$\epsilon$. This allows us to neglect the evolution of $\vec{n}_k$ and to approximate them with their
initial values. As a result we obtain that the $N$-soliton
interactions for the VNLSE in the adiabatic approximation are
modeled by the CTC, see Section 3.

It is easy to see, that if all $\langle\vec{n}_{k+1}\;^\dag,\vec{n}_k\rangle =\const \neq 0$
then the CTC (\ref{eq:141.3}) is a completely integrable dynamical system, just like the real Toda chain.

Note also that the CTC models the soliton interactions for the VNLSE with {\em any number of
components}. The effect of the polarization vectors on the interaction comes into CTC only through
the scalar products $m_{0s} =\langle\vec{n}_{k+1},\vec{n}_k\rangle $. It is well known, that
a gauge transformation $\vec{u} \to g_0 \vec{u}$ with any constant unitary matrix $g_0$ leaves the
VNLSE, Eq. (\ref{eq:vnls}) invariant.  Such transformation will change all  polarization vectors
simultaneously $\vec{n}_k \to g_0 \vec{n}_k$ but preserves their scalar products, and so
will not influence the soliton interaction.
Obviously, our CTC model is invariant under such transformations.
Due to the above arguments our choice of the initial values of $\vec{n}_{k0}$ can
be changed into $g_0\vec{n}_{k0}$ with no effect on the interaction. That is why we specify
only the scalar products $m_{k0}$ for our runs, which we have chosen to be real.

\subsection{The effects of the $\sech$-like potentials on CTC }

Now we consider the effects of the external potentials of the form (\ref{eq:iRu}). To this end we
have to calculate the integrals in the right hand side of eq. (\ref{eq:123.2}) and see how they
would change the right hand sides of eqs. (\ref{eq:139.2}).

As it is clear from above, we have to replace  $\mathcal{L}_k$ in Eq. (\ref{eq:123.2}) by
\begin{equation}\label{eq:Lkper}\begin{split}
\mathcal{L}_{k,\rm pert} = \mathcal{L}_k -2\nu_k \int_{-\infty}^{\infty}
\frac{dx \; V(x) }{\cosh^2(z_k)}.
\end{split}\end{equation}
while $\mathcal{L}_{kn}$ remains unchanged. Thus we obtain the
following PCTC system:

\begin{equation}\label{eq:Vnuk0}\begin{aligned}
{d\lambda _k \over dt} &= -4\nu _0 \left(e^{q_{k+1}-q_k} (\vec{n}_{k+1}^\dag ,\vec{n}_{k}) -
e^{q_k -q_{k-1}} (\vec{n}_{k}^\dag ,\vec{n}_{k-1}) \right) +  M_k +i N_k , \\
{d q_k \over dt} &=  - 4\nu_0 \lambda_k + 2i(\mu_0 +i\nu_0) \Xi_k  -iX_k ,\qquad
{d\vec{n}_k  \over dt } =  \mathcal{ O}(\epsilon),
\end{aligned}\end{equation}
where $\lambda _k=\mu _k+i\nu _k $,  $X_k  = 2 \mu_k \Xi_k + D_k $ and
\begin{align*}
N_k  &=- {1 \over 2} \int\limits_{-\infty}^{\infty} {dz_k \over \cosh z_k }\, \im   \left( V(y_k) u_k {\rm e}^{-i\phi_k} \right) ,
&\;  M_k  &= {1 \over 2} \int\limits_{-\infty}^{\infty} {dz_k  \, \sinh z_k \over \cosh^2 z_k }\,\re \left( V(y_k) u_k {\rm e}^{-i\phi_k} \right),\\
\Xi_k  &= -{1 \over 4 \nu_k^2} \int\limits_{-\infty}^{\infty} { dz_k \, z_k\over \cosh z_k }\,
\im\, \left(  V(y_k) u_k {\rm e}^{-i\phi_k} \right), &\; D_k & = {1 \over 2 \nu_k} \int\limits_{-\infty}^{\infty} {dz_k \, (
1 - z_k \tanh z_k)  \over \cosh z_k } \re \left(  V(y_k) u_k {\rm e}^{-i\phi_k} \right) ,
\end{align*}
where $y_k =z_k/(2\nu_0) +\xi_k$.

As a result for our specific choice of $V(x)$ we get:
\begin{equation}\label{eq:MkDk}
M_k = \sum_{s}^{} 2c_s \nu_k P(\Delta_{k,s}), \qquad N_k =0, \qquad
\Xi_k = 0, \qquad D_k = \sum_{s}^{} c_s R(\Delta_{k,s}),
\end{equation}
where $\Delta_{k,s} = 2\nu_0 \xi_k -y_s$ and
the integrals describing the interaction of the solitons with the potential (see Figure \ref{fig15}(left panel))
\begin{equation}\label{eq:P0-R0}\begin{split}
P(\Delta) &=  \frac{\Delta + 2\Delta \cosh^2(\Delta) - 3\sinh(\Delta)\cosh(\Delta) }{\sinh^4(\Delta)},\\
R(\Delta) &= \frac{6 \Delta \sinh(\Delta)\cosh(\Delta) -(2\Delta^2 +3)\sinh^2(\Delta) -3\Delta^2  }{2\sinh^4(\Delta)} .
\end{split}\end{equation}
The details of deriving the integrals are given in the Appendix.

The corrections to $N_{k} $ and $P_k$, coming from the terms
linear in $u $ depend only on the parameters of the $k $-th
soliton; \emph{i.e.}, they are `local' in $k $.

\section{CTC and the Asymptotic Regimes of $N$-soliton Trains}

The fact that the $N$-soliton trains for the scalar NLSE
are modeled by an integrable model -- CTC allowed one to
to predict their asymptotic behavior. The method to do so was based on the
exact integrability of the CTC \cite{16} and on its Lax representation.

Here we shall show, that similar results hold true also for the CTC (\ref{eq:139.2})
modeling the soliton trains of the MM.
Indeed, following Moser~\cite{36} we introduce the Lax pair
\begin{equation}\label{eq:K.3}\begin{aligned}
\dot{L} &= [B, L],
\end{aligned}\end{equation}
where
\begin{equation}\label{eq:LB}\begin{split}
L &= \sum_{k=1}^N \left( b_k E_{kk} + a_k (E_{k,k+1} + E_{k-1,k}) \right), \\
B &= \sum_{k=1}^N a_k\left( (E_{k,k+1} - E_{k-1,k} \right).
\end{split}\end{equation}
Here the matrices $(E_{kn})_{pq} = \delta_{kp}\delta_{nq} $, and
$E_{kn} = 0 $ whenever one of the indices becomes 0 or $N+1 $; the other
notations in (\ref{eq:K.3}) are as follows:
\begin{eqnarray}\label{eq:K.5}
a_k = {1 \over 2} \sqrt{\langle\vec{n}_{k+1}, \vec{n}_k\rangle} e^{(q_{k+1} - q_k)/2}, \qquad
b_k =  {1 \over 2} \left( \mu_k + i \nu_k \right).
\end{eqnarray}
One can check that the compatibility condition eq. (\ref{eq:K.3}) with $L$ and $B$ as in
(\ref{eq:LB}) is equivalent to the unperturbed CTC (\ref{eq:139.2}).

The first consequence of the Lax representation is that the CTC has $N$ complex-valued integrals of motion
provided by the eigenvalues of $L$ which we denote by  $\zeta_k =\kappa_k+i\eta_k$, $k=1,\dots,N$. Indeed the Lax equation means that
the evolution of $L$ is isospectral, \emph{i.e.}, $d \zeta_k /dt =0$.

Another important consequence from the results of Moser \cite{36} is that for the real Toda chain
one can write down explicitly its solutions in terms of the scattering data, which consist of $\{\zeta_k, r_k\}_{k=1}^{N} $ where
$r_k$ are the first components of the properly normalized eigenvectors of $L_0 $ \cite{36,44}.
For the real Toda chain both $\zeta_k=\kappa_k$ and $r_k$ are real; besides all $\zeta_k$ are different.
Next Moser  calculated the asymptotics of these solutions for $t\to\pm\infty$ and showed that $\kappa_k$
determine the asymptotic velocities of the particles.

The formulae derived by Moser can easily be extended to  the complex case \cite{13}.
The important difference is that all important ingredients such as  eigenvalues
$\zeta_k$ and first components of the eigenvectors of $L$ normalized to 1 now become
complex valued. In addition, the important asset of $L$ for the RTC, namely that
all eigenvalues are real and different, is  also lost.
However the asymptotics of the solutions for $t\to\pm\infty$ can be calculated with the
result:
\begin{equation}\label{eq:L.1}
q_k(t) = -2\nu_0 \zeta_{k} t - B_k + {\cal  O} (e^{-Dt}) ,
\end{equation}
where $D $ in (\ref{eq:L.1}) is some real positive constant which is estimated by the
minimal difference between the asymptotic velocities.
Equating the real parts in eq. (\ref{eq:L.1}) we obtain:
\begin{equation}\label{eq:xi-k}\begin{split}
\lim_{t\to\infty} (\xi_k +2\kappa_k t ) = \const
\end{split}\end{equation}
which means that the real parts $\kappa_k$ of the eigenvalues of $L$ determine the asymptotic
velocities  for the CTC.
This fact will be used to classify the regimes of asymptotic behavior.

Let us also point out the important differences between RTC and CTC, namely:

\begin{description}
  \item[D1)] While for RTC $q_k $, $r_k $ and $\zeta _k $ are all real, for CTC
they generically take complex values, e.g. $\zeta _k = \kappa _k + i
\eta_k $;

  \item[D2)] While for RTC $\zeta _k\neq \zeta _j $ for $k\neq j $, for CTC no
such restriction holds.

\end{description}

As a consequence we find that the only possible asymptotic
behavior in the RTC is the asymptotically  free motion of the
solitons. {}For CTC it is $\kappa _k $ that determines the asymptotic
velocity of the $k $-th soliton. For simplicity and without loss of
generality we assume that: $\mbox{tr}\, L_0 =0 $; $\zeta _k\neq \zeta
_j$ for $k\neq j $; and $\kappa _1 \leq \kappa_2 \leq \dots \leq
\kappa _N$. Then we have:

\begin{description}
  \item[AFR ] The asymptotically free regime takes place if
  $\kappa _k\neq \kappa _j $ for $k\neq j $, \emph{i.e.}, the
asymptotic velocities are all different. Then we have asymptotically
separating, free solitons, see also \cite{22,14};

  \item[BSR] The bound state regime takes place for $\kappa _{1} = \kappa _{2} = \dots = \kappa_{N} =0
$, \emph{i.e.}, all $N $ solitons move with the same mean asymptotic velocity,
and form a ``bound state". The key question now will be the nature of
the internal motions in such a bound state: is it quasi-equidistant or not?

  \item[MAR] a variety of intermediate situations, or mixed asymptotic regimes happen when one
group (or several groups) of particles move with the same mean
asymptotic velocity; then they would form one (or several) bound
state(s) and the rest of the particles will have free asymptotic
motion.

\end{description}

Obviously the regimes (ii) and (iii), as well as the degenerate and
singular cases, which we do not consider here have no analogies in the RTC
and physically are qualitatively different from i).

The perturbed CTC taking into account the effects of the $\sech$-like potentials to the best of our knowledge
is not integrable and does not allow Lax representation. Therefore we are applying numeric methods to solve it.

\section{Comparison Between the PCTC Model and Manakov Soliton Interactions}

In this Section we will compare how well the PCTC model derived above predicts
the soliton interactions as solutions of the MM with the external potentials of kind \eqref{eq:iRu}.
Before that let us remind the well known result (see \cite{16,22,14}) that the 3-soliton systems allow for three types of dynamical regimes
for large times, namely
\begin{description}
  \item[AFR)] asymptotically free regime when all 3 solitons move away with different velocities. This regime
  takes place if the initial amplitudes are given by eq. (\ref{eq:amp0}) with \cite{14}:
  \begin{equation}\label{eq:Deltanu}\begin{split}
  \Delta\nu < \nu_{\rm cr}=2\sqrt{2\cos(\theta_1-\theta_2)}\nu_0 \exp{(-\nu_0r_0)}
  \end{split}\end{equation}
  and the phases are as in (\ref{eq:deltak0}a).
  For our configuration with $r_0=8$ we have $\nu_{\rm cr}=0.0246$. Such asymptotic regime is shown on the left panel
  of Figure \ref{fig6};

  \item[MAR)] mixed asymptotic regime, when two of the solitons form bound state and the third soliton
  goes away from them with different velocity; Such regime takes place if the amplitudes are chosen as in (\ref{eq:Deltanu})
  and the phases are as in (\ref{eq:deltak0}b), see the left panel of Figure \ref{fig14};

  \item[BSR)] bound state regime when all solitons move asymptotically with the same velocity.
  Such regime takes place for amplitudes with $\Delta\nu >\nu_{\rm cr}$ and  the phases are as in (\ref{eq:deltak0}a).
Such asymptotic regime is shown on the left panel
  of Figure \ref{fig6a}.
\end{description}

It is natural to analyze separately all three regimes and to see what would be the effect of
the external wells/humps on them. In particular, one can determine for which positions and intensities
of the external potentials the solitons will undergo from one asymptotic regime to another.

\begin{remark}\label{rem:4}
The CTC and its perturbative version PCTC use the adiabatic approximation. If we assume that the distance between the
solitons is $r_0= 8$, then the adiabatic parameter  $\epsilon \simeq 0.01$, so one can expect that the CTC model will hold true
up to times of the order of $1/\epsilon \simeq 100$. In figure 2 we see an excellent match between the MM and CTC up to times
of the order of 300; after that the two models diverge.
\end{remark}

Since the PCTC model
is not integrable we will solve it numerically to find the predicted solitons trajectories $\xi_k(t)$.
Besides we will solve numerically the MM with the initial condition (\ref{eq:Nstr}) and
extract the trajectories of $\max(|u_1|^2 + |u_2|^2)$, where $\vec{u}\equiv(u_1, u_2)$.

\begin{figure}
\centerline{\includegraphics[width=0.5\textwidth]{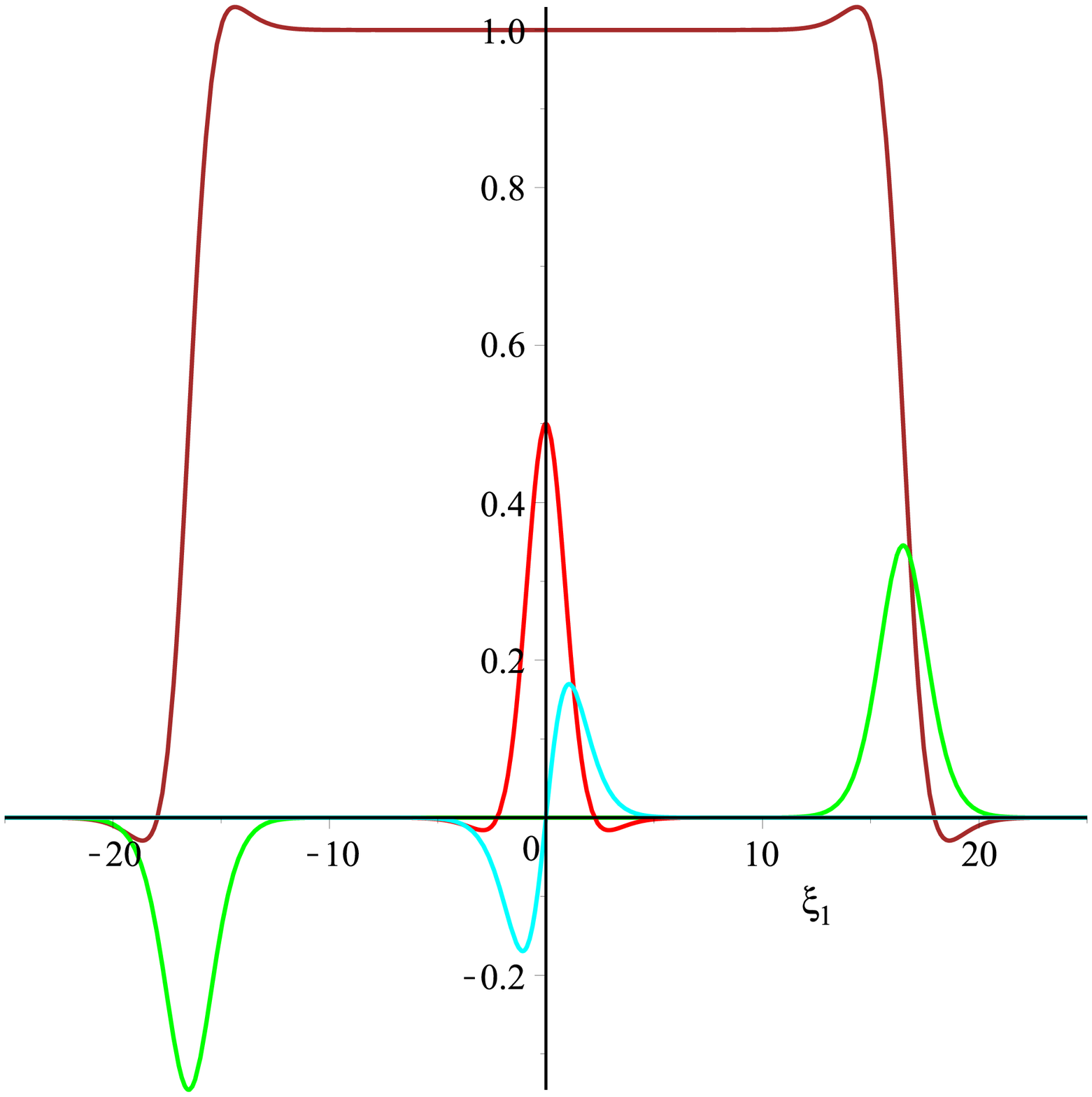}\includegraphics[width=0.5\textwidth]{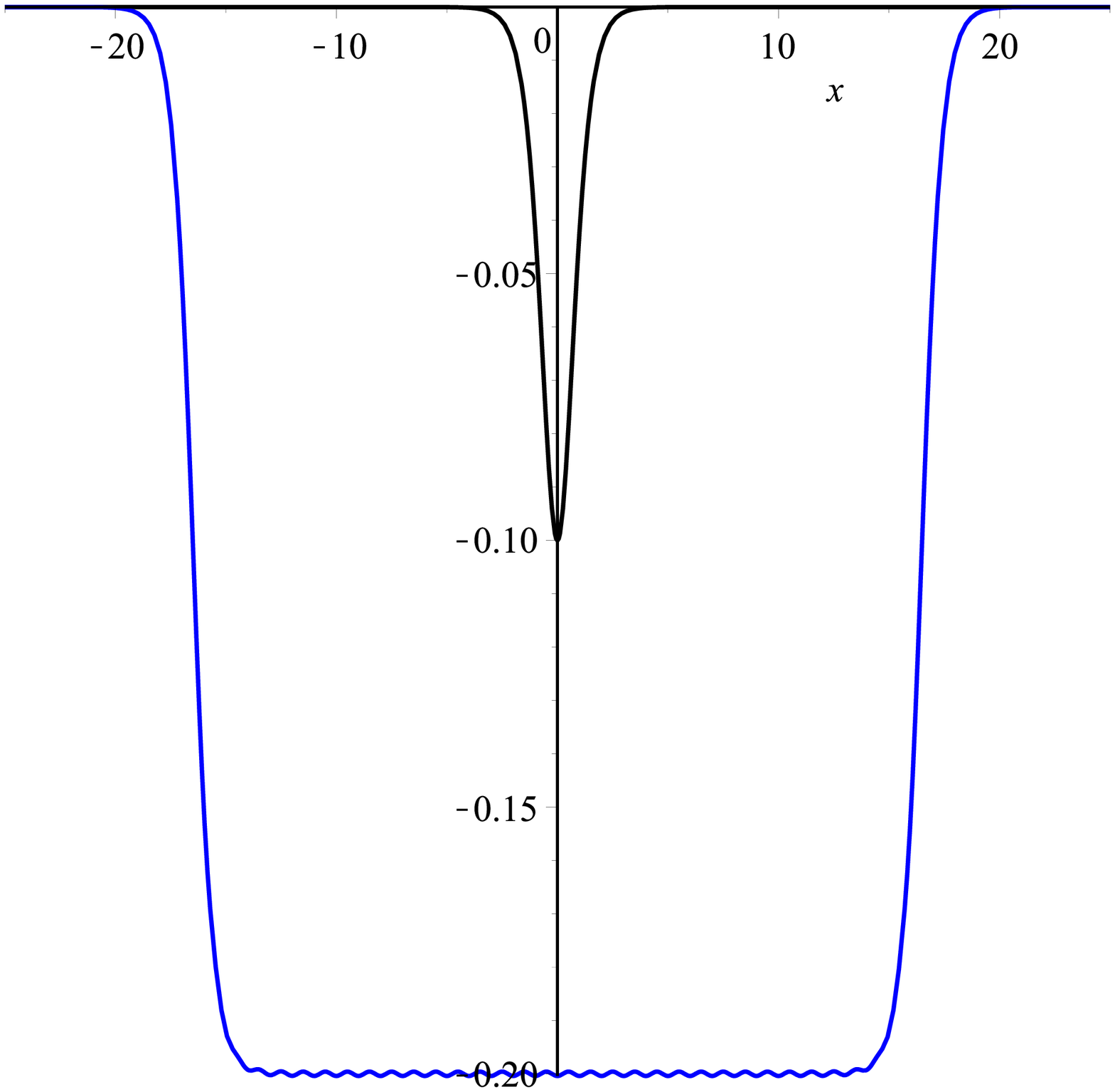}}
\vspace{-0.1in}
\caption{Graphs of $P$ and $R$ functions: for a single sech-potential centered at the origin -- in
\textit{cyan} and \textit{red} colors; and  for the superposed potential at the neighboring panel -- in \textit{green} and
\textit{brown} colors. (left); Single sech-potential in \textit{black} color vs. superposed external potential
$V(x)=\sum_{s=0}^{32} c_s {\rm sech}^2 (x-x_s)$, $c_s=-10^{-1}$, $x_s=-16+sh$, $h=1$, $s=0,...,32$ -- in \textit{blue} color.
The superposed potential forms a  well. (right).}
\label{fig15}
\end{figure}

\begin{figure}[h!]
\centerline{\includegraphics[width=0.5\textwidth]{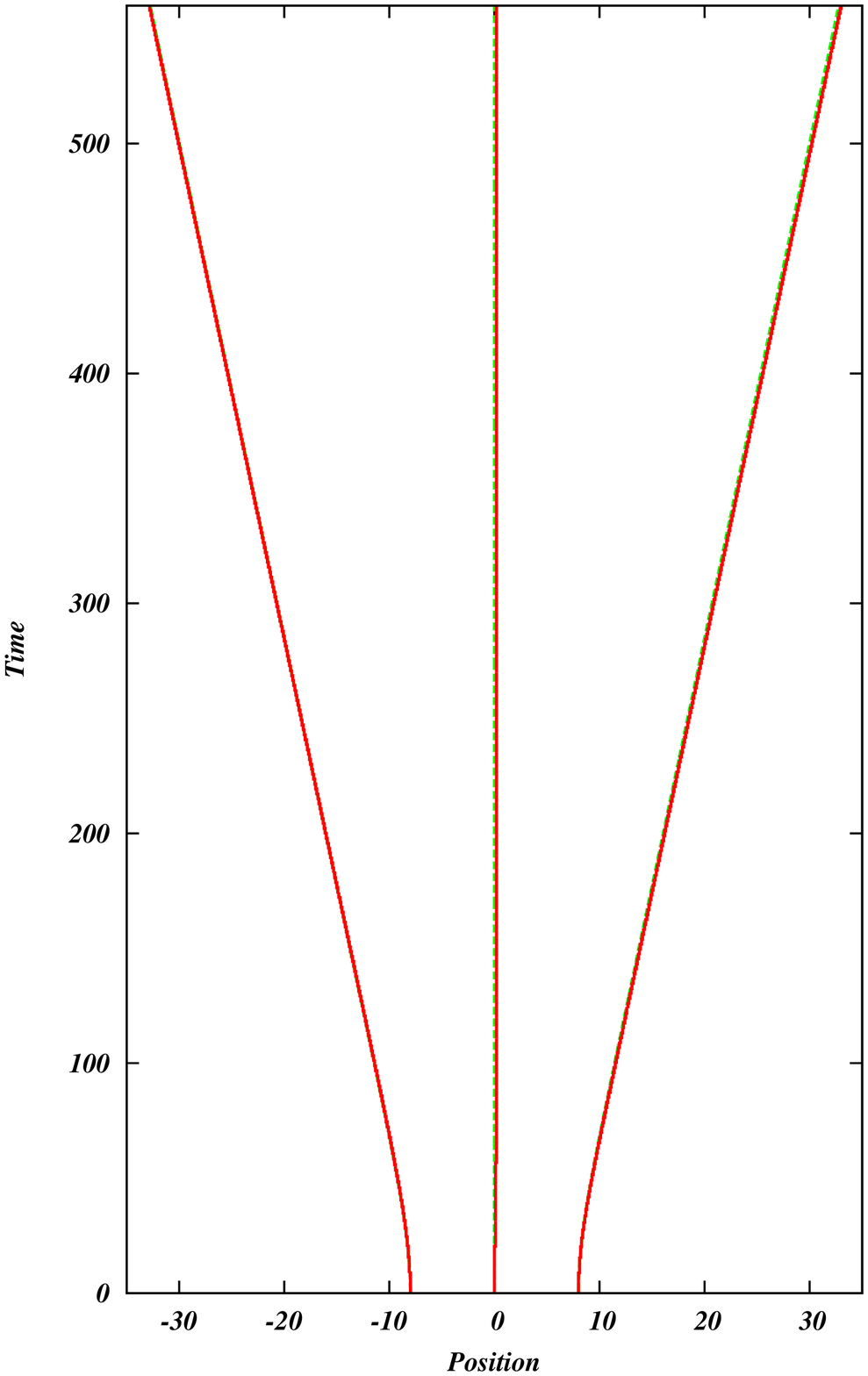}\includegraphics[width=0.5\textwidth]{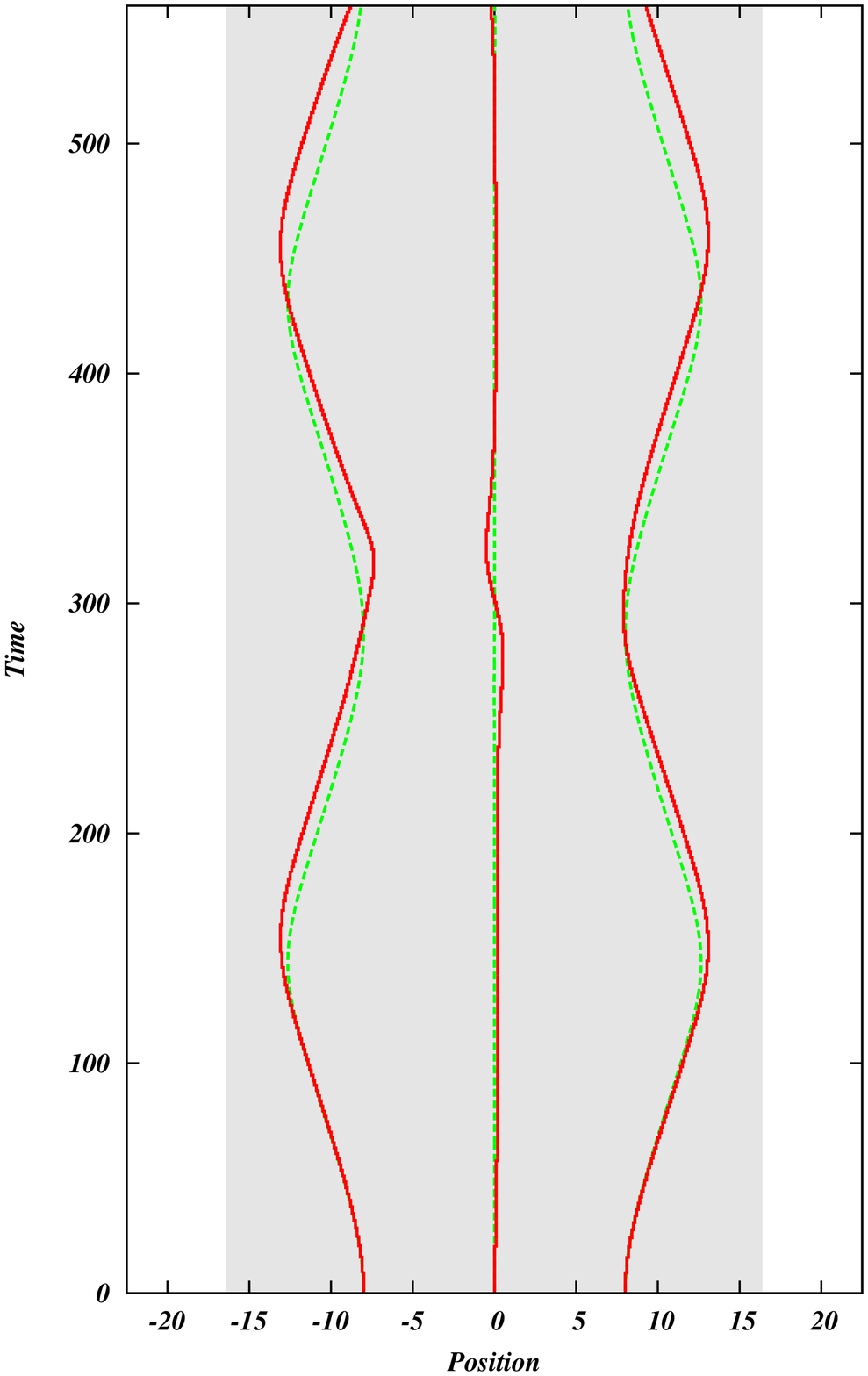}}
\vspace{-0.1in}
\caption{AFR: Free potential behavior  corresponding to real parts of eigenvalues of the Lax pair  $\re\! \zeta_1=-0.0116$, $\re\! \zeta_2=0$,
$\re\! \zeta_3=0.0116$ (left panel); External potential well $V(x)=\sum_{s=0}^{32} c_s {\rm sech}^2 (x-x_s)$, $c_s=-10^{-1}$, $x_s=-16+s$,
$s=0,...,32$. The shaded area denotes the external potential at a half level (right panel). \label{fig6}}
\end{figure}

\begin{figure}[h!]
\centerline{\includegraphics[width=0.5\textwidth]{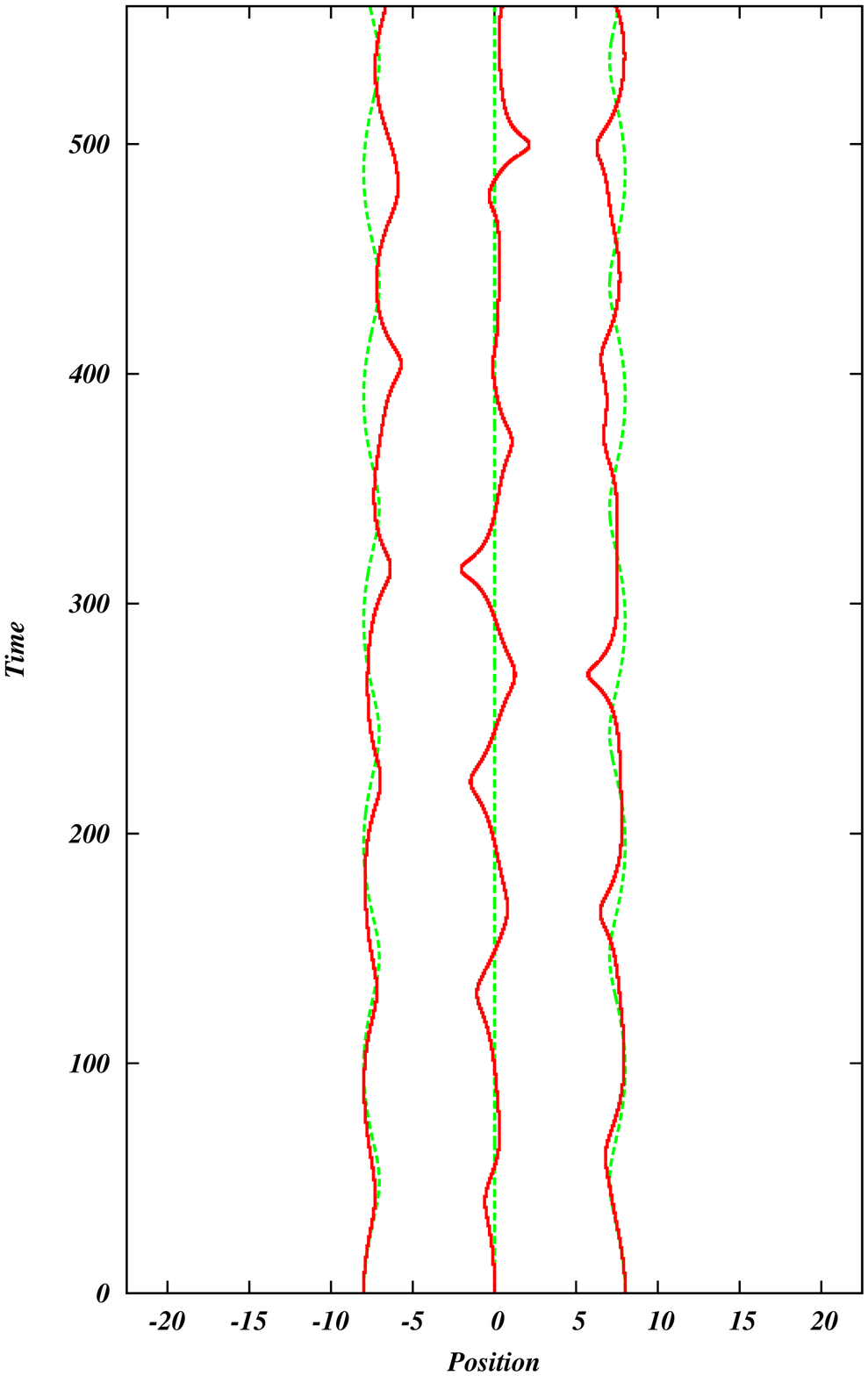}\includegraphics[width=0.5\textwidth]{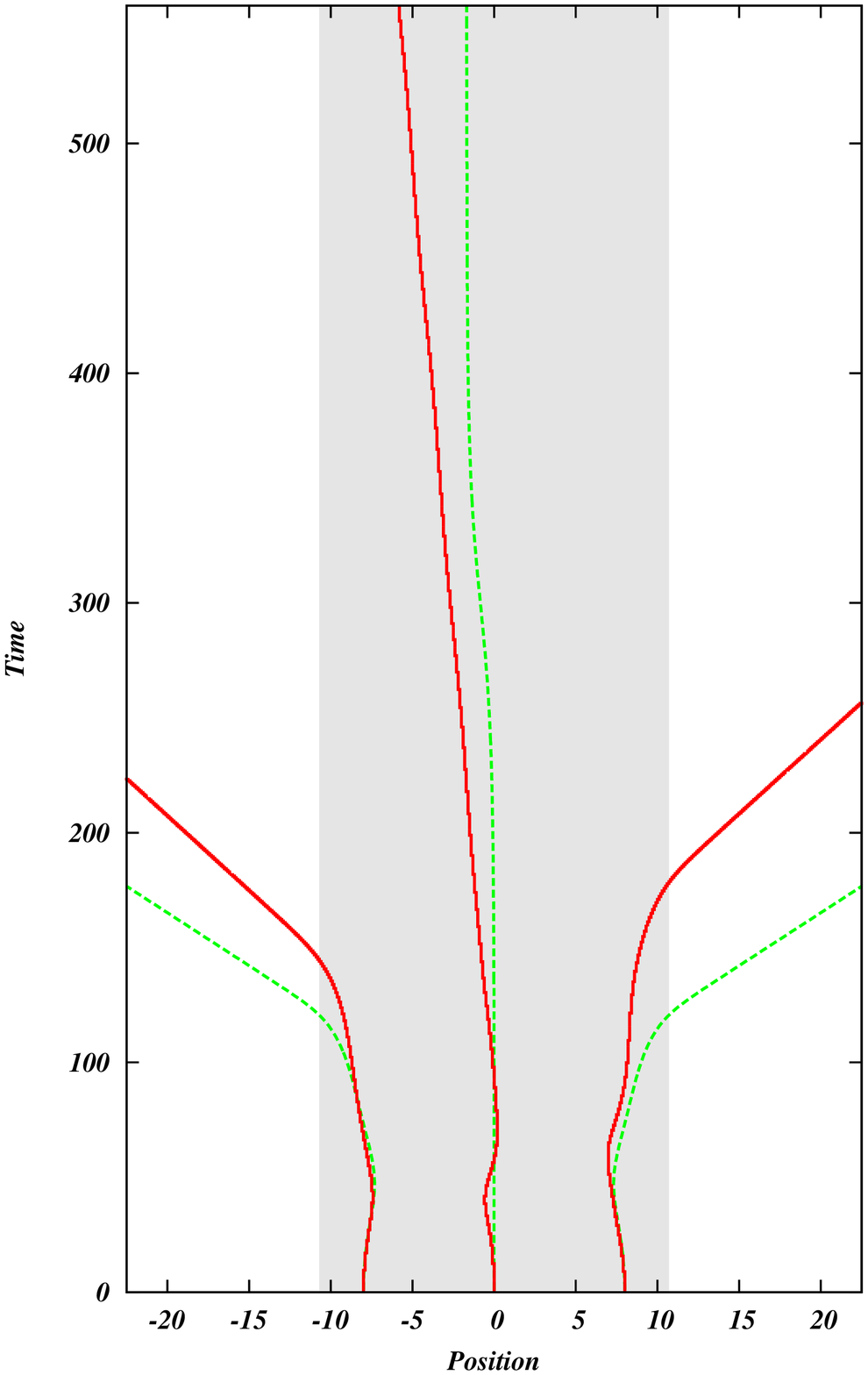}}
\vspace{-0.1in}
\caption{BSR: Free potential behavior  corresponding to real parts of eigenvalues of the Lax pair
$\re\! \zeta_1=\re\! \zeta_2=\re\!\zeta_3=0$(left); External potential hump $V(x)=\sum_{s=0}^{12} c_s
{\rm sech}^2 (x-x_s)$, $c_s=10^{-2}$, $x_s=-10+sh$, $h=5/3$, $s=0,...,12$. The shaded area denotes the external potential at a half  level (right).}
\label{fig14}
\end{figure}

On the right panel of Figure \ref{fig15} we plot samples of potential well with width 40 composed by 33 wells with depth $c_s=-0.1$
distributed uniformly between abscisas $-16$ and $16$ and distance between them $h=1$. 

Evidently each Manakov soliton solution is parameterized by 6 parameters and four of them are the usual velocity, position, amplitude and
phase. Two more parameters fix up the polarization vector. Having in mind the big parametric phenomenology of the solutions we fix the velocities,
positions and polarization vectors and vary the initial
amplitudes and phases in order to ensure one or another asymptotic regime \cite{14}.
Even with only three solitons configuration but with 13 to 33 potential wells/humps we have a large variety of combinations.

Potential wells, especially when broad enough attract the solitons and may be used to stabilize in a
bound state. Potential humps repel the solitons; choosing their positions appropriately one can either
split a soliton bound state into free solitons or force free solitons into bound state.


In what follows we compare the PCTC models with the numeric solutions of
the corresponding (perturbed) MM. In doing this, to have better base for
comparison  we keep fixed some of the the initial parameters of the soliton trains.
The other parameters may vary from run to run; their particular values will be
specified in the captions of the figures. To avoid any confusion we mark the PCTC
solutions by dashed lines, and  the numeric solutions of the MM and the perturbed MM by solid lines.
Also, we plot  the centers of solitons and track their trajectories.
Since
the PCTC are derived in the framework of the adiabatic approximation, they are expected to be adequate only up to times of the order of $\epsilon^{-1}$. So, if the distance
between neighboring solitons is 8 units, then $\epsilon \simeq 10^{-2}$ and one might expect that the PCTC
would be valid up to $t \simeq 100$. Rather surprisingly, see we find
that the models work well until $t\simeq 1000$ or even longer. We also assume that $\xi_k < \xi_{k+1}$, $k=1,2$.

Below we provide several examples that illustrate our points.
 More specifically we set:

\begin{description}
  \item[IC-1] The initial velocities $\mu_{k,0}=0$, $k=1,2,3$;

  \item[IC-2] The initial positions $\xi_{1,0} =-8, \xi_{2,0} =0,\xi_{2,0}= 8$;
  \item[IC-3] Each of the initial polarization vectors $\vec{n}_{k,0}$ will be parameterized by its
  polarization angle $\theta_{k,0}$ and a phase $\gamma_{k,0}$ as follows:
  \begin{equation}\label{eq:nk0}\begin{split}
   \vec{n}_{k,0} = \left(\begin{array}{c} {\rm e}^{i \gamma_{k,0}}\cos(\theta_{k,0}) \\  {\rm e}^{-i \gamma_{k,0}}\sin(\theta_{k,0})
    \end{array}\right).
  \end{split}\end{equation}
  Generically  the scalar products $(\vec{n}_{k+1}^\dag ,\vec{n}_{k})  $ are complex-valued. For simplicity
  here we assume that $\gamma_{k,0}=0$ for $k=1,2,3$ so that
   $(\vec{n}_{k+1}^\dag ,\vec{n}_{k}) =\cos( \theta_{k+1,0} - \theta_{k,0}) $ and $ \theta_{k,0}=(4-k)\pi/10$. Thus
   all scalar products just mentioned equal to $\cos (\pi/10) \simeq 0.951$;

  \item[IC-4] The initial amplitudes
  \begin{equation}\label{eq:amp0}\begin{split}
   \nu_{1,0}&=\nu_{0}+ \Delta \nu, \quad \nu_{2,0}=\nu_{0}=0.5, \quad \nu_{3,0}=\nu_{0}- \Delta \nu ;
  \end{split}\end{equation}

  \item[IC-5] We use two  types of initial phases configurations:
  \begin{subequations}
  \begin{align}
   &\mbox{a)} \quad \delta_{1,0} = 0, \quad  \delta_{2,0} = \pi, \quad \delta_{3,0} = 0, \quad \Delta \nu = 0.01.\notag\\
    &\hspace{0.7 cm}\text{If}\quad \Delta \nu < 0.02526 \> \text{-- asymptotically free behavior;}\label{eq:deltak0}\\
   &\mbox{b)} \quad \delta_{1,0} =\delta_{2,0} = \delta_{3,0} = 0, \quad  \Delta \nu = 0.02.\notag\\
   &\hspace{0.7 cm}\text{Bound state behavior for every}\quad \Delta \nu >0.
  \end{align}
  \end{subequations}
\end{description}

\section{Results and Discussion}
On the next figures  we show some examples of  3-soliton systems.
On the figures  we plot  the trajectories predicted by the PCTC (green dashed lines) with the MM
(red solid lines). In order to ensure the adiabaticity condition we assume that  initially the distance
between the neighboring solitons $r_0=8$.
The first example (Figure \ref{fig6}) clearly demonstrates the role of the external well on the stability of the
asymptotically free 3-soliton configuration. The potential (shaded strip) does not allow the lateral solitons to leave the
well and they start to oscillate. So, the new regime is bound state.

On the next Figure \ref{fig14} the potential free regime is bound state. The influence of potential hump of width 24
and amplitude $c_s=10^{-2}$ leads to fast violation of this regime and transition to asymptotically free behavior of the lateral solitons.

\begin{figure}[h!]
\centerline{\includegraphics[width=0.5\textwidth]{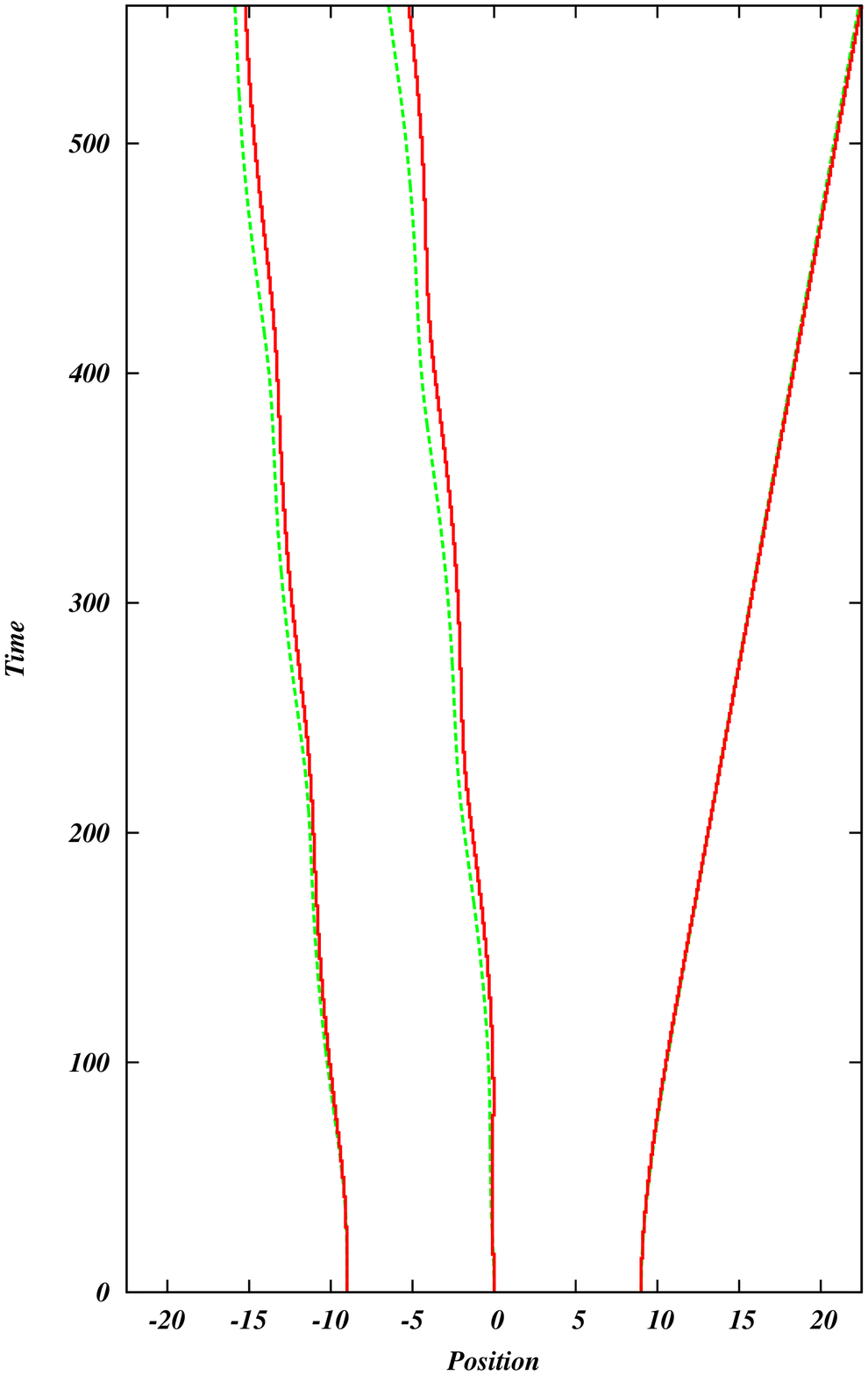}\includegraphics[width=0.5\textwidth]{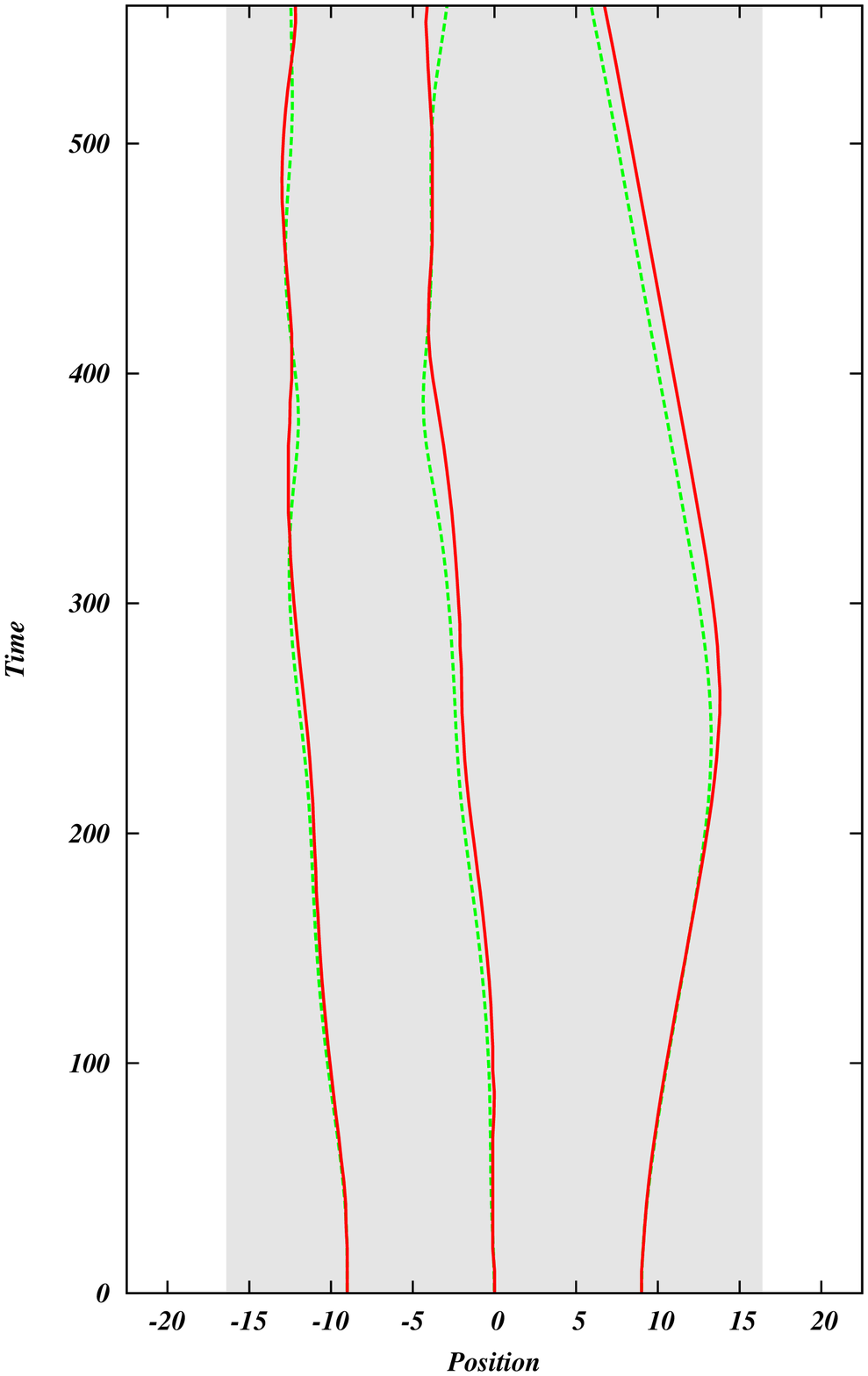}}
\vspace{-0.1in}
\caption{MAR: Free potential behavior corresponding to real parts of eigenvalues of the Lax pair
$\re\! \zeta_1=\re\! \zeta_2=-0.00321$, $\re\! \zeta_3=0.00642$ (left); External potential well
$V(x)=\sum_{s=0}^{32} c_s {\rm sech}^2 (x-x_s)$, $c_s=-10^{-2}$, $x_s=-16+sh$, $h=1$, $s=0,...,32$.
The shaded area denotes the external potential at a half level (right).}
\label{fig6a}
\end{figure}

On the Figure \ref{fig6a} is demonstrated the influence of external potential on the third possible regime -- mixed asymptotic
regime. In potential free configuration we have two bound stated solitons and one freely propagating. The adding of an external
potential as superposed wells with amplitude (depth) $c_s=-10^{-2}$ leads to a bound state behavior of all the three solitons.

 On these figures the solutions obtained by VNSE are plotted by red solid lines while those obtained by CTC model by dashed green
 line. The comparison of the numerical predictions of the both models is fully satisfactory.

\section{Conclusion}
We have analyzed the effects of the external potential wells and humps on the VNLSE soliton interactions
using the PCTC model. The comparison with the predictions of the more general VNLSE model \cite{46}
\begin{equation}\label{eq:1}
i\vec{u}_t+ \frac{1}{2} \vec{u}_{xx}+ (\vec{u}\,^\dag, \vec{u}) \vec{u}+ \alpha\vec{U}(x,t)=0,\vspace*{-0.1cm}
\end{equation}
where $\vec{U} = (|u_2|^2u_1 ,|u_1|^2u_2  )^T$ and the cross-modulation magnitude $\alpha$
is an excellent validation of the consistency and applicability of PCTC. The superposition of big number of wells/humps
obviously complexify the motion of the soliton envelopes and can cause a transition from asymptotically free and mixed asymptotic regime to a bound state
regime and vice versa. In particular, the latter means that the external potentials can be used to control the soliton
motion in a given direction and therefore to achieve a predicted  motion of the optical pulse. A general feature of the conducted
experiments is that the predictions of  both models match
very well  for a very long-time evolution. This means that PCTC is reliable model for predicting the evolution of the multisoliton
solutions of Manakov model in adiabatic approximation.

\section*{Acknowledgements}
The investigation is supported partially by Bulgarian Science Foundation under grant DDVU02/71.

These results were presented and discussed at the 8-th IMACS Conference WAVES'13 at Athens, GA and the joint seminar of The Advanced Materials Research Institute and the Department of Physics by the University of New Orleans, LA.

\appendix
\section{List of integrals}\label{sec:Int}

We outline the method of calculating the integrals \cite{21} needed to work out external
well-type potentials as in Eq. (\ref{eq:iRu}). We start with the well known integral \cite{42}:
\begin{equation*}\label{eq:Ka}\begin{split}
K(a,\Delta ) \equiv \int\limits_{-\infty }^{\infty } {dz\;{\rm e}^{iaz}  \over
2\cosh z \cosh (z+\Delta ) } = {\pi \over  \sinh \Delta} {\sin
(a\Delta/2) \over \sinh (\pi a/2) } {\rm e}^{-ia\Delta /2} .
\end{split}\end{equation*}
Using the identity $2\cosh( z -\Delta/2) \cosh (z+\Delta/2) = \cosh (2z) +\cosh(\Delta)$ and
applying the substitution $z\to z-\Delta/2$ and we get
\begin{equation*}\label{eq:KKa}\begin{split}
{\rm e}^{i a \Delta/2} K(a,\Delta) = \int\limits_{-\infty}^{\infty} \frac{dz {\rm e}^{iaz}}{\cosh (2z) + \cosh (\Delta)} =
\frac{\pi \sin (a\Delta/2) }{\sinh(\Delta) \sinh(\pi a/2)}.
\end{split}\end{equation*}

Next we differentiate with respect to $\Delta$ and divide by $\sinh \Delta$:
\begin{equation*}\label{eq:KK2a}\begin{split}
\int\limits_{-\infty}^{\infty} \frac{dz {\rm e}^{iaz}}{(\cosh (2z) + \cosh (\Delta))^2} =
 - \frac{1}{\sinh(\Delta)}\frac{\partial }{ \partial \Delta} \left(
\frac{\pi \sin (a\Delta/2) }{\sinh(\Delta) \sinh(\pi a/2)}\right),
\end{split}\end{equation*}
\begin{equation*}\label{eq:KK3a}\begin{split}
\int\limits_{-\infty}^{\infty} \frac{dz {\rm e}^{iaz}}{(\cosh (2z) + \cosh (\Delta))^3} =
\frac{1}{\sinh(\Delta)}\frac{\partial }{ \partial \Delta} \left( \frac{1}{\sinh(\Delta)}\frac{\partial }{ \partial \Delta} \left(
\frac{\pi \sin (a\Delta/2) }{\sinh(\Delta) \sinh(\pi a/2)}\right) \right).
\end{split}\end{equation*}
Thus we have:
\begin{equation*}\label{eq:N-R2}\begin{split}
N(a,\Delta) &= - \frac{2}{\sinh(\Delta)} {\rm e}^{-ia\Delta/2} \frac{\partial }{ \partial \Delta }
\left( {\rm e}^{ia\Delta/2} K(a,\Delta) \right),\qquad
P(\Delta) = - \left. \frac{1}{2}{\rm e}^{ia\Delta} \frac{\partial N(a,\Delta) }{ \partial \Delta } \right|_{a=0}, \\[-0.2cm]\\
R(\Delta)&=  N(0,\Delta) - \frac{1}{2i} \left. \frac{\partial }{ \partial a } \left(
{\rm e}^{ia\Delta} \frac{\partial N(a,\Delta)}{ \partial \Delta } \right)\right|_{a=0} .
\end{split}\end{equation*}

If we have generic (complex-valued) potential $V(x)$ we will have to calculate the integrals:
\begin{equation*}\label{eq:N-R}\begin{aligned}
N(\Delta) &= \int\limits_{-\infty}^{\infty} \frac{dz }{2 \cosh^2(z) \cosh^2(z-\Delta )}= \frac{2\Delta\cosh(\Delta) - 2\sinh(\Delta) }{\sinh^3(\Delta)},\\
Q(\Delta) &= \int\limits_{-\infty}^{\infty} \frac{dz z}{2 \cosh^2(z) \cosh^2(z-\Delta )}
=\frac{2\Delta\sinh(\Delta) - 2\Delta^2\cosh(\Delta) }{\sinh^3(\Delta)},
\end{aligned}\end{equation*}
while $P(\Delta)$ and $R(\Delta)$ are given in eq. (\ref{eq:P0-R0}).

\end{document}